\documentclass[runningheads]{llncs} 

\usepackage[utf8]{inputenc}
\usepackage{algorithm}
\usepackage[]{algpseudocode}
\usepackage{hyperref}
\usepackage{gensymb}
\usepackage{subcaption}
\usepackage{comment}
\usepackage{amsmath}
\usepackage{todonotes}
\usepackage{xspace}

\usepackage[misc]{ifsym}

\usepackage{amsmath,amssymb,color}
\usepackage[normalem]{ulem}
\usepackage{xcolor}

\title{Stress-Plus-X (SPX) Graph Layout}
\author{Sabin Devkota(\Letter)\orcidID{0000-0002-0610-6573}, Reyan Ahmed\orcidID{0000-0001-6830-9053}, Felice De Luca\orcidID{0000-0001-5937-7636}, Katherine~E.~Isaacs\orcidID{0000-0002-9947-928X}, Stephen~Kobourov\orcidID{0000-0002-0477-2724}}

\institute{Department of Computer Science, University of Arizona, USA\\
\email{\{devkotasabin, abureyanahmed, felicedeluca\}@email.arizona.edu,\\ \{kisaacs, kobourov\}@cs.arizona.edu}
 }

\begin{document}
	
	\maketitle
	\begin{abstract}
		
		Stress, edge crossings, and 
		crossing angles play an important role in the quality and readability of graph drawings. 
		Most standard graph drawing algorithms optimize one of these criteria which may lead to layouts that are deficient in other criteria. We introduce an optimization framework, Stress-Plus-X (SPX), that simultaneously optimizes stress together with several other criteria: edge crossings, minimum crossing angle, and upwardness (for directed acyclic graphs). SPX 
		achieves results that are close to the state-of-the-art algorithms that optimize these metrics 
		individually. SPX is flexible and extensible and can optimize a subset or all of these criteria simultaneously. Our experimental 
		analysis shows that our joint optimization approach is successful in drawing graphs
		with good performance across readability criteria.
		
	\end{abstract}

\section{Introduction}

\let\thefootnote\relax\footnote{$^1\text{This work is supported in part by NSF grants CCF-1740858, CCF-1712119 and}\\ \text{ DMS-1839274, DMS-1839307.}$}


Several criteria have been proposed for evaluating the quality of graph
layouts~\cite{ware2002cognitive}, including minimizing stress, minimizing the number of edge crossings, minimizing drawing area, as well as maximizing the angle between edge crossings, maintaining
separation between marks (``resolution''), and preserving
highly connected neighborhoods.  In the case of directed acyclic graphs (DAGs), maintaining consistent edge direction, i.e., upwardness, is preferable. While these criteria  have been shown to improve human performance for graph tasks, automatic layout approaches actively target at 
most one from the list.

We propose a framework, \textit{Stress-Plus-X} (\textit{SPX}), for automatic layout of node-link
diagrams that targets multiple graph layout criteria simultaneously.
SPX formulates the layout as an optimization problem that combines stress minimization with penalty
 terms representing other criteria. Composing and weighting the terms in the objective function provides the flexibility and extensibility.

\begin{figure}[htp]
	\centering
	\includegraphics[width=1.0\textwidth]{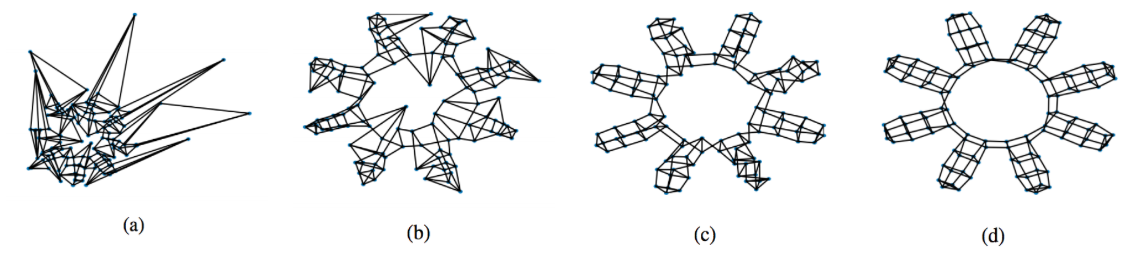}
	\caption{Different layouts of the same graph from the crossing angle maximization Graph Drawing Contest: (a) from the T\"ubingen algorithm that won in 2018~\cite{Bekos18}; (b) from the KIT algorithm that won in 2017~\cite{Demel2018AGH}; (c-d) from SPX  with different balance in the optimization of stress, crossing angle, and edge crossings.}
	\label{figure:input_graphs}
\end{figure}

\begin{figure}[!ht]
	\centering
	\includegraphics[width=.9\textwidth]{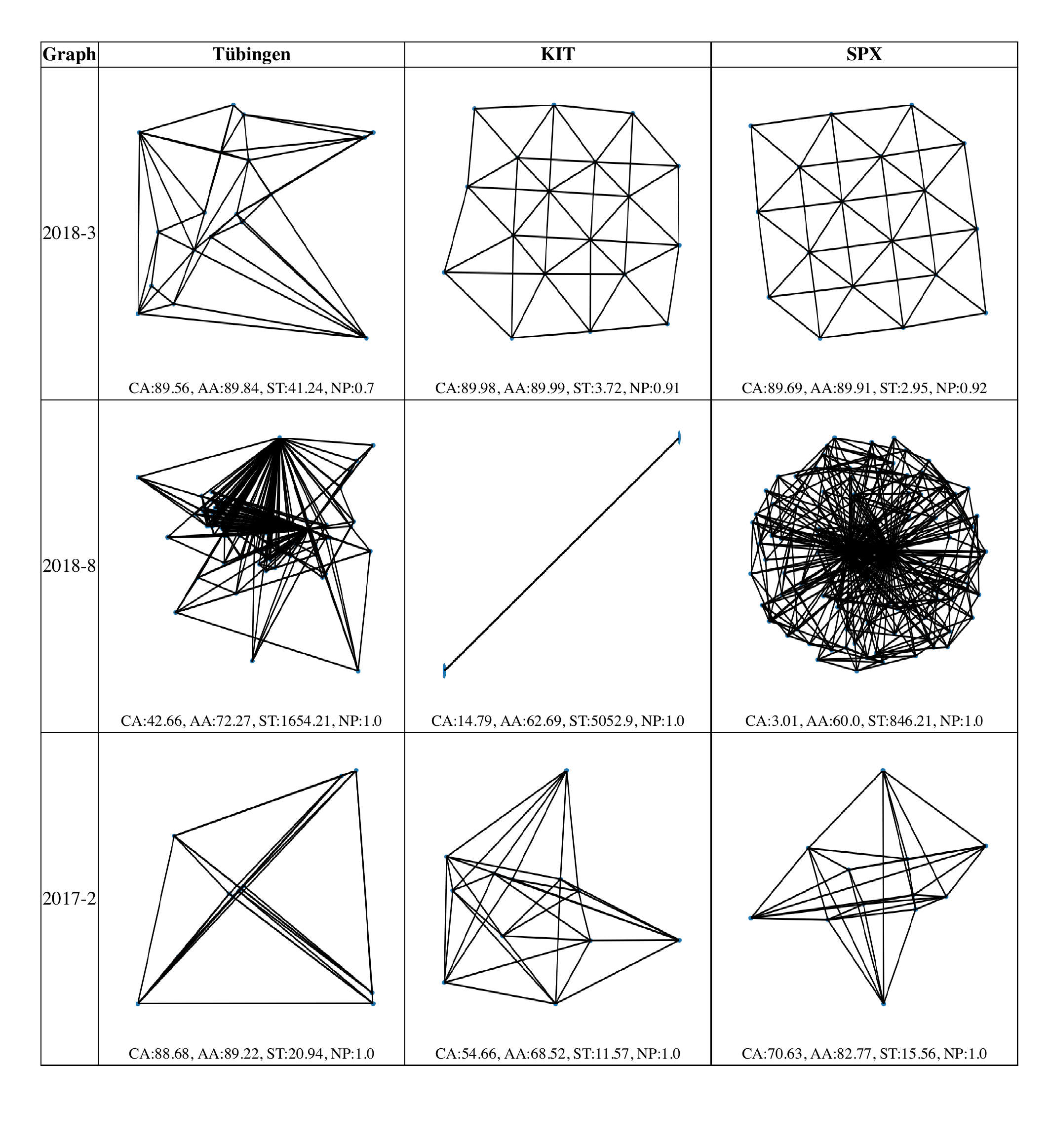}
	\caption{
	Graphs from the 2017-18 Graph Drawing Contests. In graph 2018-3, the crossing angles are all within 1\% of the optimal, yet SPX best shows the underlying graph structure. 
	The best crossing angle layout for Graph 2018-8 (center) yields a
large drawing area.
	The best crossing angle layout for Graph 2017-2 (left) yields poor
vertex resolution. We report crossing angle (CA), average angle (AA), stress (ST), and neighborhood preservation (NP).}
	\label{fig:comparison_1_4}
\end{figure}

With the adage \textit{``Don't let perfect be the enemy of good''} in mind, the
goal of SPX is not to optimize any one particular criterion at the cost of all others, but to find a balance across the criteria as optimizing only one criterion can lead to poor quality drawings~\cite{huang2013}.
As an extreme example, for minimum drawing area we can place all vertices on top of each other, yet perform poorly in the other quality criteria. A similar example is shown in Figure~\ref{figure:input_graphs} where (a-b) show the outputs on a Graph Drawing Contest graph produced by two state-of-the-art algorithms for crossing angle maximization~\cite{Bekos18,Demel2018AGH} while (c-d) show the outputs of SPX with different balance in the optimization of stress, crossing angle, and edge crossings. Note that the SPX approach better preserves topology and produces visually appealing results and although (d) has the lowest crossing angle, it arguably provides the most recognizable drawing.
Delving further into this observation, we examined the contest graphs across several metrics, as shown in Figure~\ref{fig:comparison_1_4}, noting that optimizing for one criterion could yield extreme drawings. Graph 2018-8 in the middle row is a case where optimal crossing angle (center) requires a very large drawing area. Graph 2017-2 in the last row is a case where the best crossing angle (left) exhibits poor vertex resolution. These observations motivated us to seek a balance of criteria to improve drawings.


To demonstrate our framework, we formulate optimization terms for three
criteria: minimizing edge crossings, maximizing the crossing angle, and upwardness (all of which have been used in Graph Drawing Contests).
We 
compare our edge crossing formulation to  state-of-the-art approaches 
on a corpus of community graphs. SPX
achieves better edge crossing results than just optimizing stress, and frequently
outperforms several of the crossing-centric algorithms. 
Similarly, we show that aiming only at the optimization of crossing angle tends to significantly impact the quality of the layout for other criteria. 
Although the angle-centric algorithms outperform SPX, our algorithm generates comparable crossing angle values and sometimes outperforms the angle-centric ones, while still achieving better
performance on other drawing aspects.   Finally, we compare our upwardness preserving approach to existing directed graph layout approaches~\cite{dagrejs,ipsepcola_2006,gansner1990technique,mutzel2007ogdf}. 


In summary, our contributions are: (1) Stress-Plus-X (SPX), a framework for optimizing multiple graph drawing	criteria simultaneously (Section~\ref{sec:SPX}); (2) Optimization terms for maximizing edge crossing angles (Section~\ref{subsec:angle}) and upwardness preservation (Section~\ref{subsec:upward}); and (3) An evaluation of our optimization terms in comparison to
	state-of-the-art single criterion approaches (Section~\ref{sec:results}).

\section{Background and Related Work}\label{sec:background}

Existing graph layout algorithms usually optimize a single drawing criterion, e.g., minimizing stress or maximizing the minimum edge crossing angle. We define these criteria formally and discuss layout approaches
that focus on them.

\paragraph{Stress:}  Stress measures the difference between node-pair distances in a layout and their graph-theoretic distances, based on an all-pairs shortest path computation. It is a natural measure of how well the layout captures the structure in the underlying graph.
Let $\textbf{C}_i$ be the position of the $i$th node in a layout $\textbf{C}$ and $d_{ij}$ be the graph distance between node pair $i,j$. Then 
stress$(\textbf{C}) = \sum_{i<j} (w_{ij} ||\textbf{C}_i - \textbf{C}_j|| - d_{ij})^2$. A typical normalization value is $w_{ij} = d^{-2}_{ij}$.

Kamada and Kawai~\cite{kamada_1989} formulate the graph layout problem as that of minimizing 
stress and use energy-based optimization.
Gansner \textit{et al.}~\cite{stress_maj_2005} use stress majorization instead. 
Stress-based graph visualization can also be seen as a special case of a multi-dimensional scaling (MDS)~\cite{kruskal1964multidimensional,shepard1962analysis}, which is a powerful dimensionality reduction technique.
Variants of MDS are used in many graph layout systems, including~\cite{chen2009local,stress_maj_2005,pich2009applications}. None of these methods aim to optimize other criteria such as minimizing edge crossings or maximizing crossing angles.

Wang \textit{et al.}~\cite{revisited_stress_2018} reformulate stress to incorporate target edge directions and lengths and propose constraints to reduce crossings or improve crossing angle in given subgraphs, but not in the entire graph.
Constrained layout algorithms~\cite{scalable_cola_2009,ipsepcola_2006} combine stress minimization or force-directed layout with separation constraints between node pairs. Constrained layouts, however, do not optimize for  edge crossings or crossing angles. When used with force-directed layout algorithms (such as Fruchterman-Reingold~\cite{fr_reingold_1991}) instead of with stress minimization, stress is also not optimized.

\paragraph{Edge Crossings and Crossing Angles:} Minimizing the number of crossings between edges in
a graph layout has been shown to be an important heuristic in readability of
graphs~\cite{Purchase1997}, prompting interest in several graph drawing
contests~\cite{Abrego12,Buchheim13}. 
Other than recent works by Radermacher \textit{et al.}~\cite{Radermacher18} and Shabbeer \textit{et al.}~\cite{bennett2010} (discussed in Section~\ref{subsec:related-joint}), there is little work on directly minimizing edge crossings in general graphs. 


The crossing angle of a straight-line drawing of a graph is the smallest angle between two crossing edges in the layout. Large crossing angles have been shown~\cite{Argyriou2010,huang2014,huang2013} to improve graph readability and 
several heuristics have been proposed to maximize crossing angles.
Demel \textit{et al.} (KIT)~\cite{Demel2018AGH} propose a greedy heuristic to select the best
position for a single vertex from a random set of points. 
Bekos \textit{et al.} (T\"ubingen)~\cite{Bekos18} propose selecting a vertex arbitrarily from a set
of vertices, called the \textit{vertex-pool}, which contains a subset of the
vertices which are adjacent to the pairs of edges that have the minimum
crossing angle. 
Both approaches above performed very well
in crossing angle maximization, but neither is concerned with stress minimization or other criteria.

\paragraph{Upward Drawing:} A drawing of a directed acyclic graph is upward if the target vertex of each directed edge has a strictly higher $y$-coordinate than the source vertex. Upward drawing is used to show ordering or precedence between entities in a variety of settings
~\cite{ipsepcola_2006,gansner1990technique}.
 Sugiyama layout~\cite{sugiyama1981methods} is the most common approach for creating upward drawings. The layout algorithm assigns ranks to the vertices to determine their $y$-coordinates followed by computing their $x$-coordinates to minimize crossings between consecutive layers. Examples include \textit{dot}~\cite{gansner1990technique}, \textit{dagre}~\cite{dagrejs}, and \textit{OGDF}~\cite{mutzel2007ogdf}. Mixed graphs, where only subgraphs are drawn upward, have also been drawn using this approach~\cite{seemann1997extending}.
 

\paragraph{Neighborhood preservation and Drawing Area:}
While  stress  captures  how  well  {\em global}  graph  distances  are  realized  in  the layout, neighborhood preservation captures how well {\em local} neighborhoods are preserved in the layout. This is the optimization goal of more recent dimensionality reduction techniques such as t-SNE~\cite{maaten2008visualizing} and UMap~\cite{mcinnes2018umap}.
Specifically, in the context of graph drawing, neighborhood preservation is defined as the Jaccard similarity between the adjacent nodes in the graph and the nearest nodes in the layout, averaged over all nodes in the graph~\cite{tsne_2017}. 

Drawing area refers to the size of the canvas used to layout the graph and is implicit when nodes are placed on an integer grid. Large drawing area is undesirable due to difficulties navigating the visualization or resolving the marks. Minimizing drawing area has also been used in Graph Drawing Contest challenges~\cite{duncan2012graph,gutwenger2014graph}.

\paragraph{Joint Optimization:}
\label{subsec:related-joint}

Our work aims to jointly optimize several graph drawing heuristics simultaneously. Huang \textit{et al.}~\cite{huang2013} previously optimized for two criteria simultaneously, namely crossing angle and angular resolution of the graph in a force-directed setting. Shabbeer \textit{et al.}~\cite{bennett2010} minimized stress and edge crossings simultaneously using an optimization-based approach. 


The objective function of Shabbeer \textit{et al.} contains penalties for edge crossings. Edge crossings can be expressed as a system of non-linear constraints. Consider two edges $\textbf{A} = \bigl(\begin{matrix}
a_1^x & a_1^y \\ a_2^x & a_2^y
\end{matrix} \bigr)$ and $\textbf{B} = \bigl(\begin{matrix}
b_1^x & b_1^y \\ b_2^x & b_2^y
\end{matrix} \bigr)$ where the two nodes of $\textbf{A}$ are $(a_1^x , a_1^y)$, and $(a_2^x , a_2^y)$ and similarly for $\textbf{B}$.
Farkas' Theorem can be used to state that the edges $\textbf{A}$ and $\textbf{B}$ do not cross if and only if there exists 
$\textbf{u}$, and $\gamma$, such that


\begin{equation}
\label{eq2}
\textbf{Au} + \gamma \textbf{e} \geq \textbf{0}, \textbf{Bu} + (1+\gamma) \textbf{e} \leq \textbf{0}
\end{equation}
where $\textbf{e}$ is a 2-dimensional vector of ones. Intuitively, Eq.\ref{eq2} states that for a pair of edges $\textbf{A}$ and $\textbf{B}$ to not cross, there must exist a line that strictly separates the edges $A$ and $B$, i.e., there is a non-zero margin between them. Here, $\textbf{u}$ refers to a vector that is perpendicular to the direction of the separating line and 
$\gamma$ is a scalar value that ensures the non-zero margin of separation between the edges. 

This set of inequalities can be transformed into a penalty term, $penalty(\textbf{A}, \textbf{B}),$ for edge pair $\textbf{A}, \textbf{B}$ such that it is zero for non-crossing edge pairs and strictly positive for crossing edge pairs.

\begin{equation}
\label{eq3}
penalty(\textbf{A}, \textbf{B}) =  \min\limits_{u,v}||(-\textbf{Au} -\gamma \textbf{e})_+||_1 + ||(\textbf{Bu} + (1+\gamma)\textbf{e})_+||_1
\end{equation}
where  $(z)_+ = max(0, z)$. The penalty term is combined with stress as a cost function and then iterative optimization is used to compute a layout. They demonstrate their approach on small biological networks.

Our approach differs in that our goal is a framework for balancing multiple criteria to achieve good results across them. We introduce penalties and constraints for crossing angle maximization and upward drawings. We further introduce a weighting to the edge crossings. Finally, we introduce a hyperparameter to directly balance across criteria.


\section{SPX Algorithm}
\label{sec:SPX}

Stress-Plus-X (SPX) is a unified framework that can simultaneously optimize stress along with other graph drawing criteria. The ``X'' in SPX refers to the constraints that encode the additional criteria. We describe cost functions for encoding the number of edge crossings and crossing angle respectively, as well as constraints for preserving upwardness. The general SPX model is as follows:
\begin{align}
cost(\textbf{C},\textbf{u},\gamma, \boldsymbol{\rho}) = stress(\textbf{C}) + K \times \sum Penalties(\textbf{C},\textbf{u},\gamma, \textbf{P})
\end{align} 
with node coordinates \textbf{C}, balancing hyperparameter $K$, optional penalty parameters $P$ (e.g., $\rho_i$ in Section~\ref{subsec:spcrossings}), and $\gamma$ and \textbf{u} as described in Section~\ref{subsec:related-joint}. 

Intuitively, decreasing stress, decreasing the penalty term for X, or decreasing both results in a decrease in the objective function.
Hence, minimizing the objective function simultaneously optimizes for both stress and ``X."

Modifying the value of $K$ allows us to control the balance between the stress and the ``X'' terms. Figure~\ref{figure:input_graphs} (c) and (d) show two layouts of the same graph created with different $K$ parameterizations. Adjusting $K$ to better balance criteria can result in a more intuitive drawing.


\subsubsection{Optimization Procedure}
We optimize the cost function iteratively in two phases. We first compute the optimal $\textbf{u}$ and $\gamma$ for each pair of edges $(\textbf{A}, \textbf{B})$ via linear programming to minimize the penalties, $penalty(\textbf{A},\textbf{B})$. Then, keeping the $\textbf{u}'s$ and $\gamma$'s constant, for all edge pairs, we optimize the cost function by modifying $\textbf{C}$ using gradient descent; see Algorithm~\ref{alg:spx}.

\begin{algorithm}
	\caption{Stress-plux-X($G$)}
	\begin{algorithmic}
		\State Compute initial layout $\textbf{C}_0$ (using stress majorization, force-directed layout, or random initialization)
		\For{Number-of-iterations}
		\State Keeping the node coordinates $\textbf{C}$ constant, find optimal $\textbf{u}$ and $\gamma$ for each edge pair (\textbf{A},\textbf{B}) using linear programming to minimize $penalty(\textbf{A},\textbf{B})$
		\State Keeping $\textbf{u}$'s and $\gamma$'s constant, minimize $cost(\textbf{C},\textbf{U},\gamma, \boldsymbol{\rho}) $ by updating $\textbf{C}$ using gradient descent
		\EndFor

	\end{algorithmic}
	\label{alg:spx}
\end{algorithm}

\subsection{Stress plus Crossing Minimization}
\label{subsec:spcrossings}

The edge crossings penalty is: $\sum_{i=1}^{l} (\rho_i/2) * \{|| (-\textbf{A}^{i}(\textbf{C})u^i - \gamma^{i}\textbf{e})_{+}  ||_1 \nonumber + ||(\textbf{B}^i(\textbf{C})u^i  + (1+\gamma^i)\textbf{e} )_{+}||_1\}$
where $l$ is the number of edge pairs, $\textbf{A}^i(\textbf{C})$ and $\textbf{B}^i(\textbf{C})$ are the first and second edges of edge pair $i$ as matrices $\textbf{A}$ and $\textbf{B}$, $\textbf{u}^i, \gamma^i$ are the $\textbf{u},\gamma$ terms for edge pair $i$, and $\rho_i$ is a weight on the penalty for edge pair $i$.

Shabbeer \textit{et al.}~\cite{bennett2010} use a compounding weight where each edge crossing gets penalized more the longer it persists through the optimization iterations. We found that such a penalty can result in the introduction of new edge crossings for graphs that are larger and denser. With this in mind, we use a binary weight for $\rho_i$: the value is $1$ when edges intersect and $0$ otherwise.  

The cost function for stress plus crossing minimization further differs from Shabbeer \textit{et al.} in the criteria weighting parameter $K$.  Figure~\ref{fig:min_ncr_100_part} (Section~\ref{sec:results}) shows that the use of binary weights and hyperparameter $K$ helps  SPX achieve better results compared to Shabbeer \textit{et al.}. 

\subsection{Stress plus Crossing Angle Maximization}
\label{subsec:angle}

Our crossing angle maximization penalty is the edge crossing penalty with an additional factor of $cos^2(\theta_i)$ in each factor of the summation, where $\theta_i$ is the angle between a pair of crossing edges. We use $cos^2$ to constrain to positive values and give a heavier weight to smaller crossing angles. Note this modified penalty function explicitly maximizes the minimum crossing angle and implicitly minimizes the number of crossings, as when a crossings is removed altogether it cannot contribute to the minimum crossing angle.

\subsection{Stress plus Upward Crossing Minimization}
\label{subsec:upward}

We add the upwardness criteria to SPX by adding constraints to the model. Let $(u, v)$ be a directed edge. Then, in the drawing of the graph the $y$ coordinate of $v$ should be strictly larger than the $y$ coordinate of $u$. We enforce this directly with a linear constraint ($y_v > y_u$). If the input graph is a DAG then we add this constraint for all edges. If the graph is mixed then we add the constraints only for the directed edges.

\subsection{Implementation}
\label{sec:implementation}




We implemented SPX in Python. It uses the stress majorization formulation of Gansner \textit{et al.}~\cite{stress_maj_2005} to minimize stress and the edge crossing detection code from Demel\textit{ et al.}~\cite{Demel2018AGH}.
SPX source code and experimental material are available at \url{https://github.com/devkotasabin/SPX-graph-layout}.

\subsubsection{Initial Layouts}
We ran our experiments using 3 different layout algorithms as input to the SPX algorithm: stress majorization (\texttt{neato}), force-directed layout (\texttt{sfdp}), and random initialization. Both \texttt{neato} and \texttt{sfdp} are available in the GraphViz package~\cite{ellson2001graphviz}. To ameliorate the effects of sensitivity to initial layout, we employ random starts of SPX, using each method multiple times and choosing the layout that maximizes the objective. 

\subsubsection{Gradient Descent Algorithms}
\label{subsec:gd}
We experimented with the following algorithms for gradient descent (GD)~\cite{gd_overview_2016}: \texttt{bfgs}, \texttt{l-bfgs}, vanilla GD, momentum-based GD, Nesterov momentum-based GD, \texttt{Adagrad}, \texttt{RMSprop}, and \texttt{Adam}.
We found that for different types of graphs, different GD variants yielded better results and we kept all but \texttt{bfgs} and \texttt{l-bfgs} in our parameter sweep based on their performance in our pilot experiments. Section~\ref{sec:convergence} contains further analysis of different GD variants and their convergence plots.

	\subsubsection{Parallelization}
	Each combination of random initial layout, gradient descent algorithm, and value of $K$ is independent and thus can be run in parallel. Operations on each edge pair, such as computing $\textbf{u}$ and $\gamma$, as well as summing the penalties, can also be parallelized. However, running edge pairs fully in parallel would incur significant overhead. We leave the implementation of this approach as future work.

\subsection{Convergence analysis}
\label{sec:convergence}

\begin{figure}[htp]
		\centering
		\fbox{
		\begin{subfigure}[b]{0.32\textwidth}
			\includegraphics[width=\textwidth]{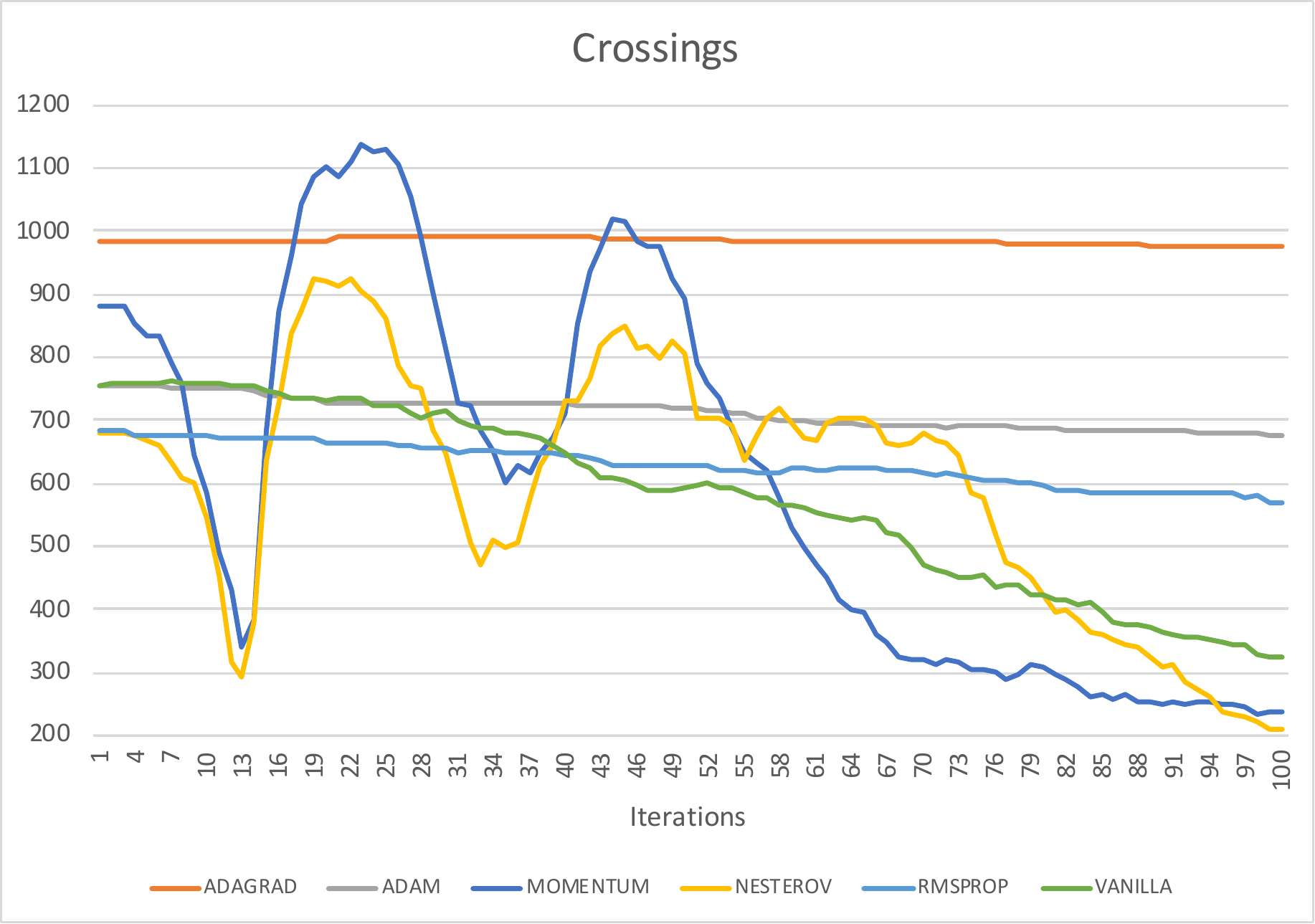}
			\caption*{} 
		\end{subfigure}
		\begin{subfigure}[b]{0.32\textwidth}
			\includegraphics[width=\textwidth]{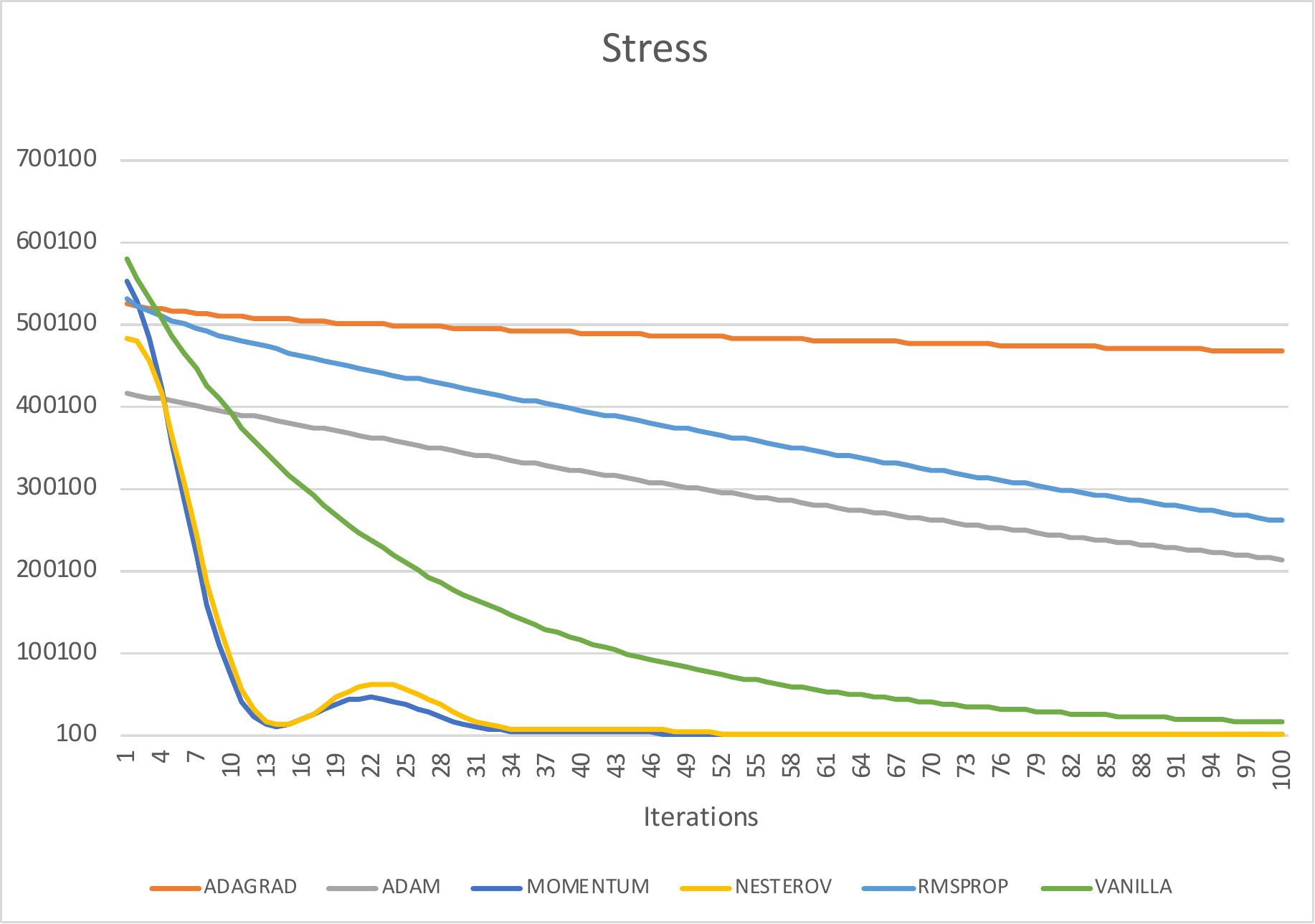}
        \caption*{Community graph 5} 
		\end{subfigure}
		\begin{subfigure}[b]{0.32\textwidth}
			\includegraphics[width=\textwidth]{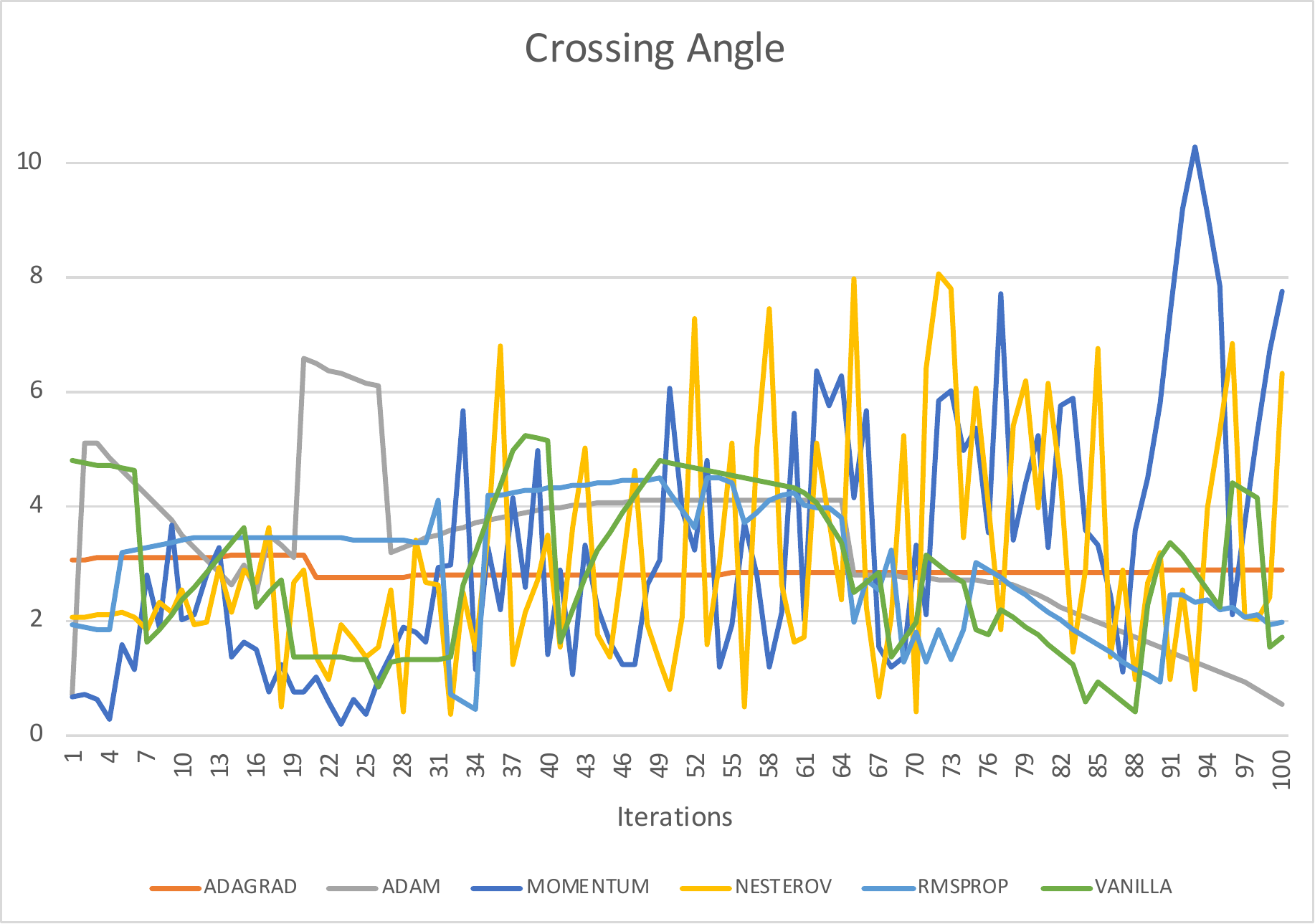}
			\caption*{} 
		\end{subfigure}}\\
		\fbox{
		\begin{subfigure}[b]{0.32\textwidth}
			\includegraphics[width=\textwidth]{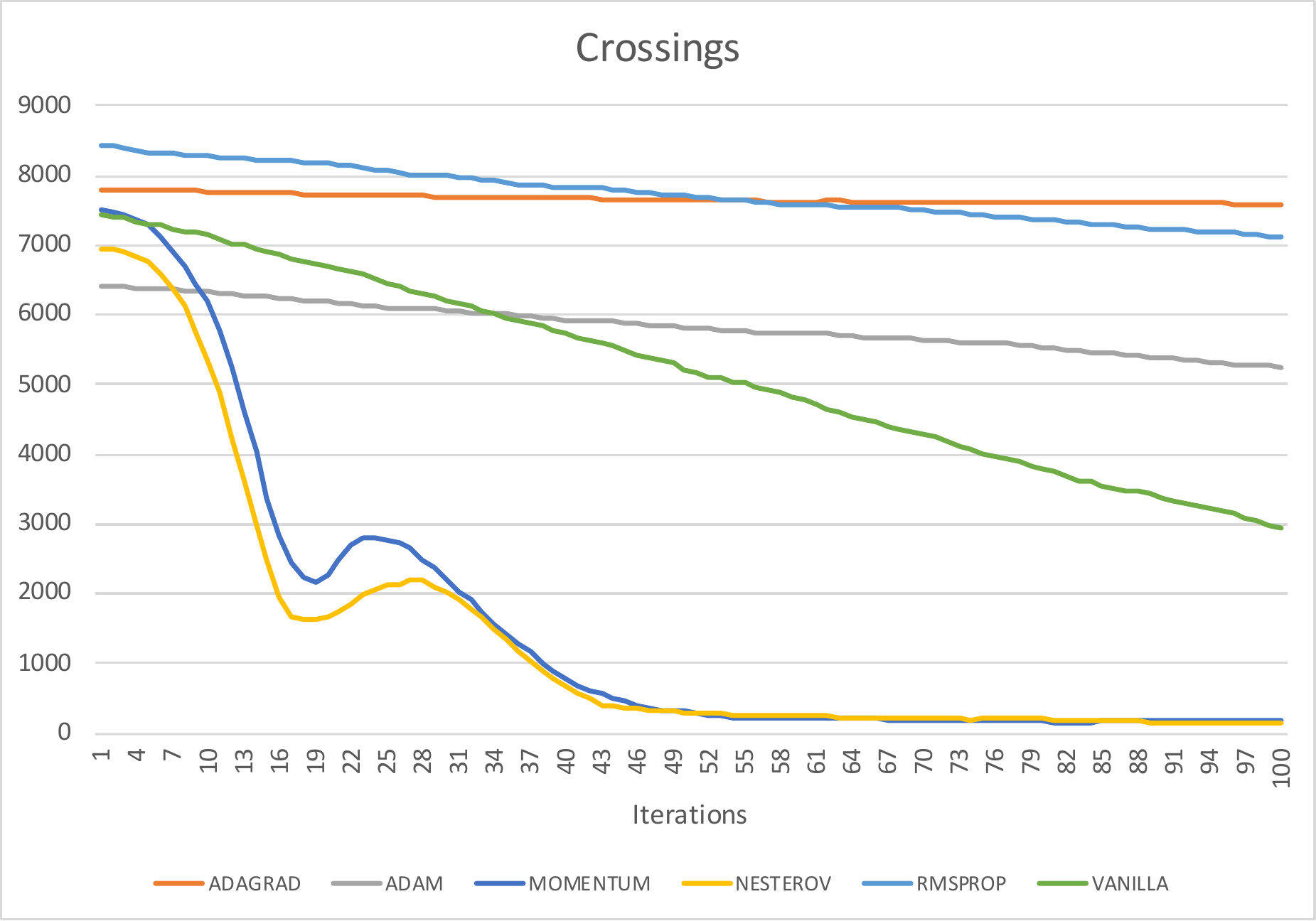}
			\caption *{} 
		\end{subfigure}
		\begin{subfigure}[b]{0.32\textwidth}
			\includegraphics[width=\textwidth]{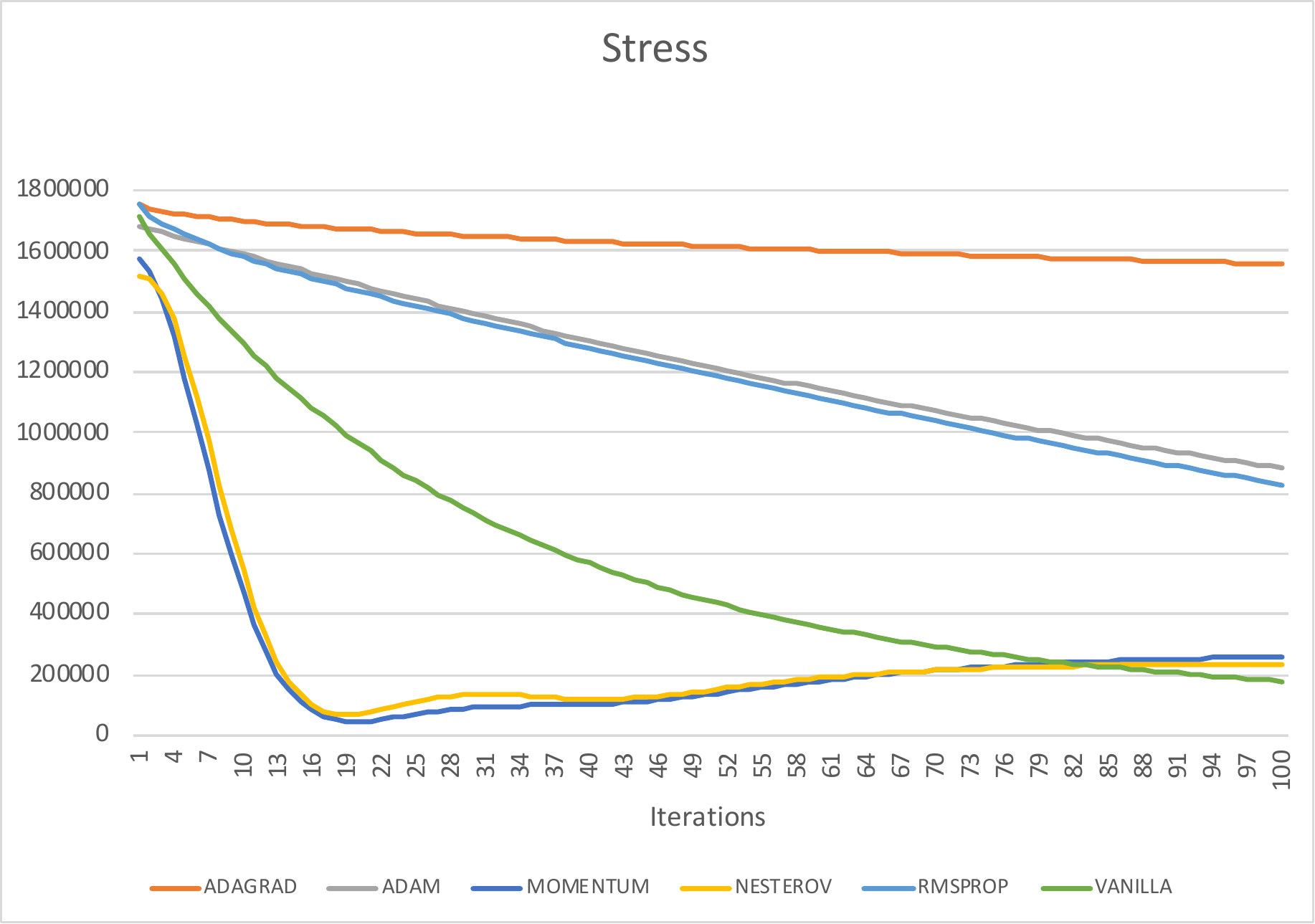}
			\caption*{Contest graph 9 from 2018} 
		\end{subfigure}
		\begin{subfigure}[b]{0.32\textwidth}
			\includegraphics[width=\textwidth]{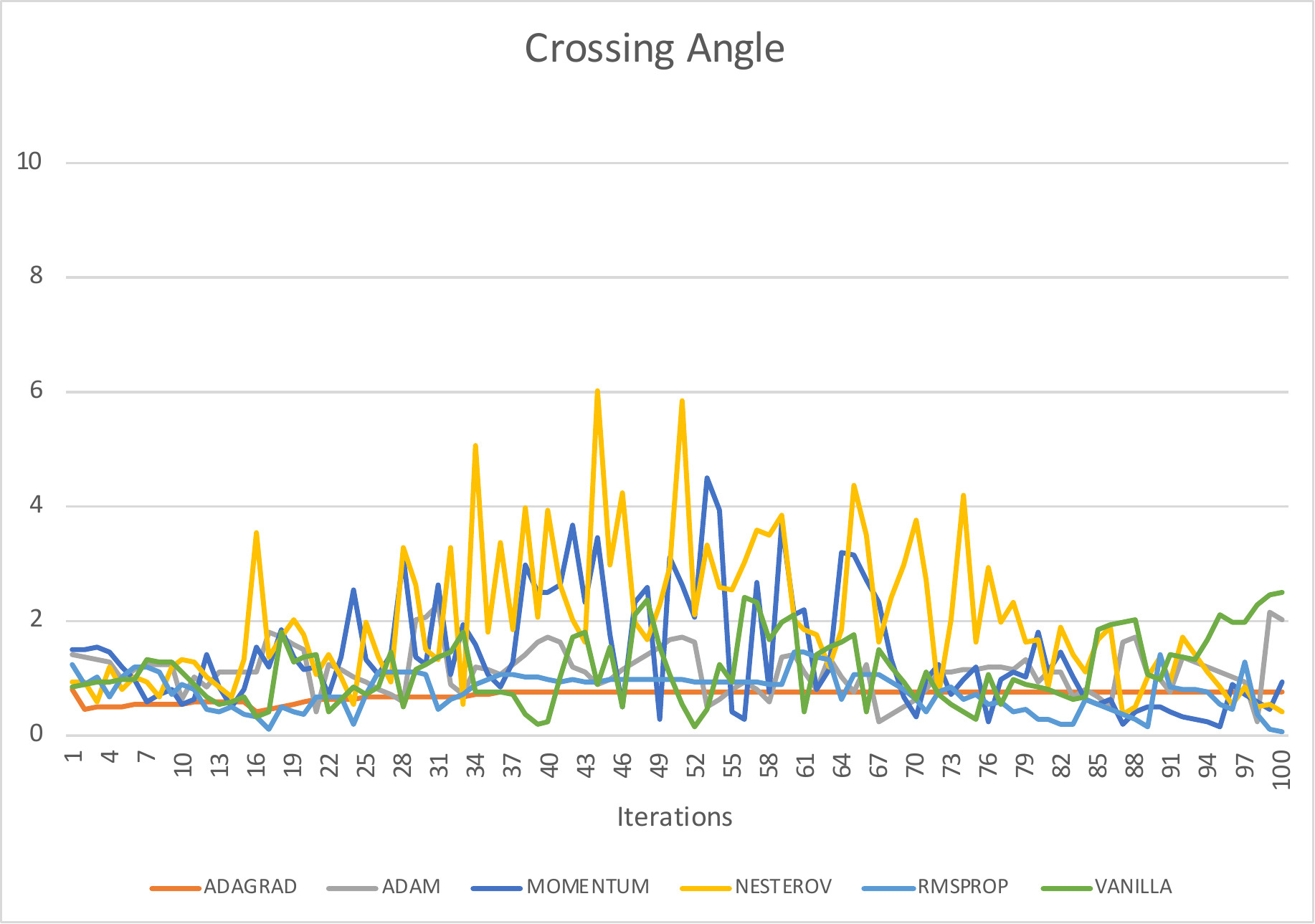}
 			\caption*{} 
		\end{subfigure}}%
		\caption{Number of crossings, stress, and crossing angle over 100 iterations for 6 variants of GD algorithms on 2 graphs run with fixed K$(=2)$. The 2 graphs are graph 5 from random subset of 25 community graphs (top row) and graph 9 from 2018 graph drawing contest (bottom row).}
		\label{fig:GDtrends}
	\end{figure}

Figure~\ref{fig:GDtrends} illustrates the convergence behavior of SPX using the six variants of gradient descent from Section~\ref{sec:implementation} on two graphs,  graph 5 from the community graphs of Section~\ref{sec:results} (top row) and graph 9 from 2018 Graph Drawing contest (bottom row). Convergence behavior of the variants differ depending on graph. Figure~\ref{fig:GDtrends} shows the values for number of crossings, stress, and crossing angle over 100 iterations for a fixed value of K$(=2)$ for both graphs.

	For both graphs, at least one gradient descent variant converges within 100 iterations. 
	In the first graph, \texttt{Momentum} and \texttt{Nesterov} converge rapidly and then get stuck in local minima. In the first graph, they overcome the local minima to continue convergence, while on the second graph they diverge after the minima. We hypothesize convergence per variant is dependent on graph properties and thus use all six. Further analysis and optimization is left for future work.
	
	
\section{Results}
\label{sec:results}

As SPX is designed to be a flexible framework, we evaluate it in three different contexts. First, we compare SPX to Shabbeer \textit{et al.}~\cite{bennett2010} on stress and number of crossings showing SPX performs better.

Second, we compare SPX to two state-of-the-art algorithms for crossing angle optimization: Demel \textit{et al.} from KIT ~\cite{Demel2018AGH} and Bekos \textit{et al.} from T\"ubingen~\cite{Bekos18}. 
We compare across five readability metrics discussed earlier: stress (ST), number of crossings (NC), crossing angle (CA), drawing area (DA), and neighborhood preservation (NP). We show SPX balanced multiple criteria simultaneously rather than optimizing one at the expense of others.



Third, we compare SPX to existing approaches that directly optimize crossings~\cite{dagrejs,gansner1990technique,mutzel2007ogdf} for upward drawings of DAGs. Our results show that SPX can preserve upwardness while performing better across other readability criteria.

\subsection{Datasets and Experimental Settings}

For the first two evaluations we used the 2017 and 2018 graph drawing contest graphs~\cite{devanny2017graph,devanny2018graph}, as well as a collection of 400 graphs used in a crossings minimization study by Radermacher \textit{et al.}~\cite{Radermacher18}. In this paper we discuss results on different subsets of these datasets, and more details are provided in the appendix.
For the third evaluation (upward drawing of DAGs) we generated 4 trees and 30 DAGs of different sizes.

We ran our experiments using all six gradient descent variants discussed in Section~\ref{subsec:gd}. We swept the values of $K$ in the range of $2^{-5}$ to $2^5$ in exponential increments. We used three different initial layout algorithms as input: \texttt{neato}, \texttt{sfdp}, and random initialization with five different starts each. Metrics were calculated using the \texttt{graphmetrics} library of De Luca~\cite{felicemetrics}.

\subsection{Comparison to Shabbeer \textit{et al.}}
We compare SPX with two algorithms - Shabbeer \textit{et al.}~\cite{bennett2010} and stress majorization~\cite{stress_maj_2005} on the corpus of 100 community graphs. The crossings value for stress is taken from Radermacher \textit{et al.}~\cite{Radermacher18} and the stress value calculated as lowest from five random \texttt{neato}~\cite{ellson2001graphviz} layouts. We run the SPX variant that performs stress-plus-crossing minimization only and compare using two metrics, number of crossings and stress, because Shabbeer \textit{et al.} minimizes only for these two metrics. We do not perform the same two-metric comparison with the crossing minimization algorithms of Radermacher \textit{et al.}~\cite{Radermacher18} because they are not concerned with stress. We provide details about crossing minimization only for SPX and the algorithms of Radermacher \textit{et al.} in the appendix.

Figure~\ref{figure:min_ncr_100_part_crossing}
shows that on average SPX produces fewer crossings than both other approaches. Figure~\ref{figure:min_ncr_100_part_stress} shows that on average SPX produces layouts with lower stress than both other approaches. We hypothesize that SPX performs better than stress majorization on stress because of SPX's multiple random starts and the use of \texttt{neato} as one of the initializations.

\begin{figure}[htp]
	\centering
	\begin{subfigure}[b]{0.3\textwidth}
		\includegraphics[width=\textwidth]{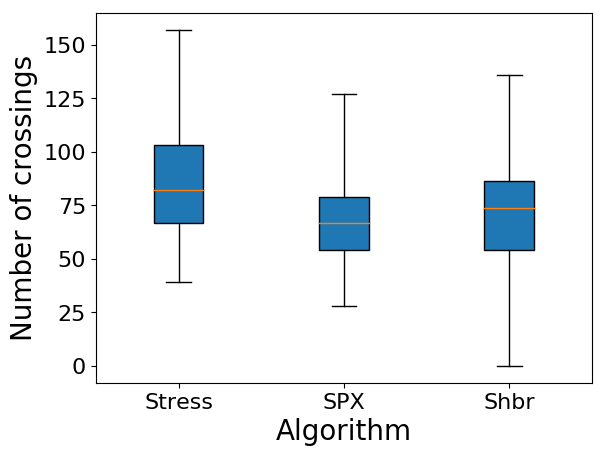}
		\caption{Crossing minimization} \label{figure:min_ncr_100_part_crossing}
	\end{subfigure}
	~
	\begin{subfigure}[b]{0.3\textwidth}
		\includegraphics[width=\textwidth]{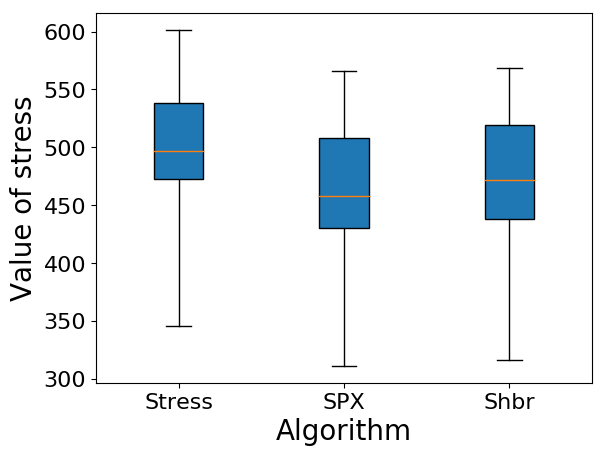}
		\caption{Stress minimization} \label{figure:min_ncr_100_part_stress}
	\end{subfigure}
	\caption{Comparing SPX, Shabbeer \textit{et al.} and stress majorization in terms of the number of crossing and stress minimization using 100 community graphs.}
	\label{fig:min_ncr_100_part}
\end{figure}

\subsection{Comparison Across Several Criteria}

We examine several readability criteria across the layouts obtained by the three algorithms designed to minimize crossing angle: KIT~\cite{Demel2018AGH}, T\"ubingen~\cite{Bekos18}, and SPX. In particular, we consider stress, neighborhood preservation, edge crossings, drawing area, and crossing angle.  


Though our impetus was the graph drawing contest graphs, they are diverse in structure, making it difficult to compare across them. To perform a bulk comparison, we randomly select a subset of 25 graphs from community graphs described above.

\begin{figure}[!ht]
	\centering
	\includegraphics[width=.8\textwidth]{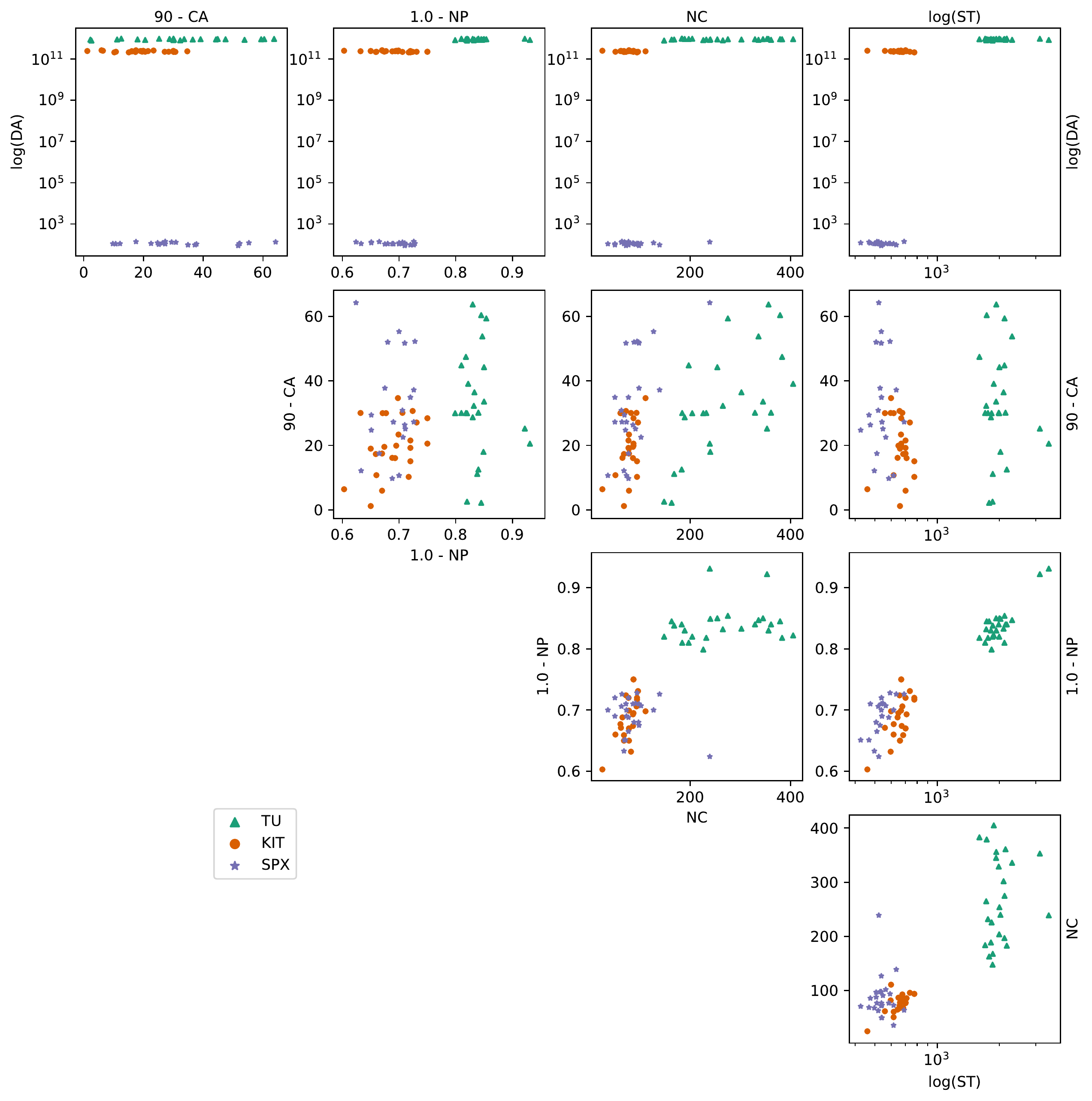}
	\caption{Pairwise metric evaluation of the KIT, T\"ubingen, and SPX algorithms using stress (ST), number of crossings (NC), crossing angle (CA), neighborhood preservation (NP), and drawing area (DA). 
	} 
	\label{figure:metric_comparison_plot}
\end{figure}

Figure~\ref{figure:metric_comparison_plot} shows the results for the 25 graphs, presented in a pairwise fashion of metrics. We plot the metrics so that points in the lower left corner indicate
good performance in the two metrics. From the plots we can see that most of the SPX drawings are in the well-performing corner. 

\begin{figure}[!ht]
	\centering
	\includegraphics[width=0.7\textwidth]{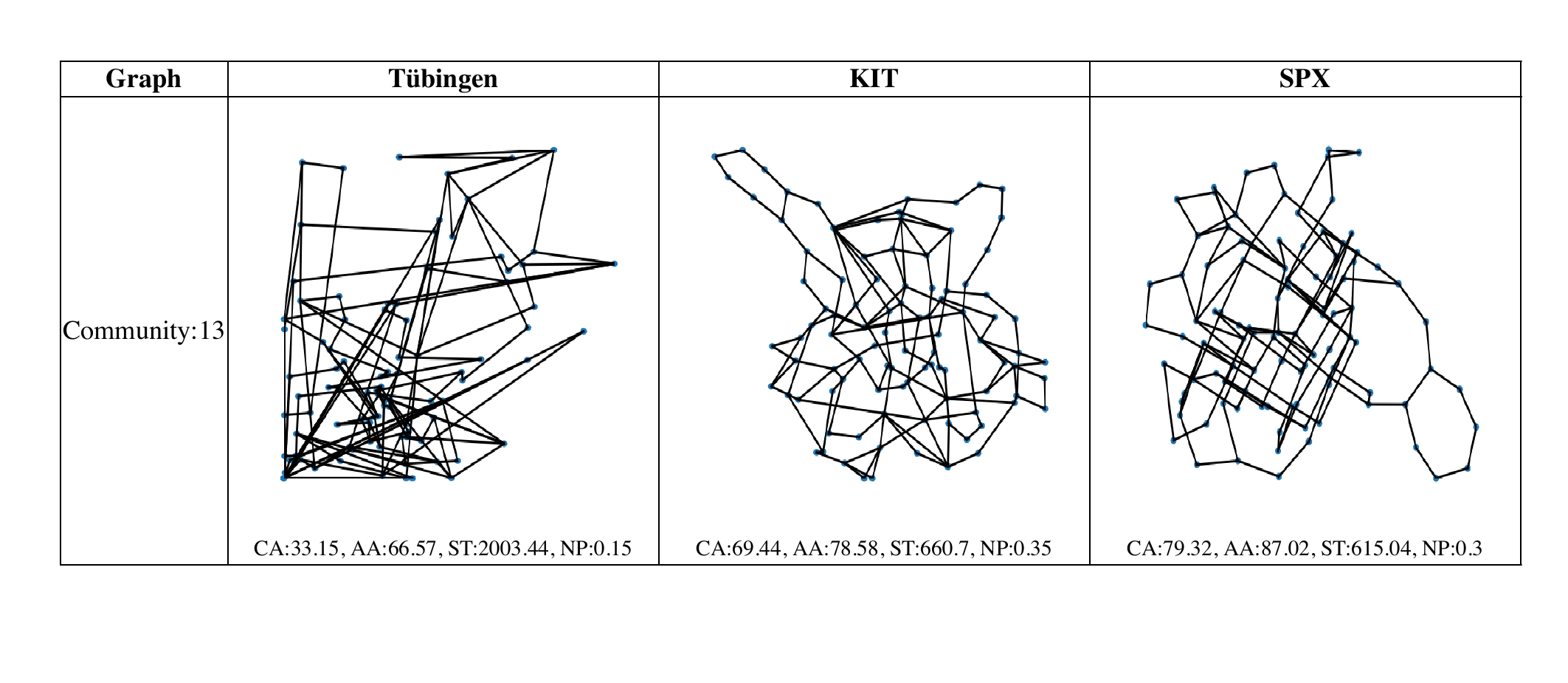}
	\caption{
	Outputs of the T\"ubingen, KIT, SPX algorithms on a community graph.
	}
	\label{figure:community_graph_summary_single}
\end{figure}

Figure~\ref{figure:community_graph_summary_single} shows an example of a community graph, drawn by all three algorithms. SPX achieves best stress and crossing angle while performing very close to the winner, KIT, in terms of number of crossings. 


\subsection{Comparison of Upward Drawings}
    
    To evaluate SPX for upward drawing, we compare it to several state-of-the-art directed graph algorithms across several metrics on a corpus of 4 trees and 30 DAGs, described in Appendix~\ref{sec:dags}.
    
    We compared SPX to \textit{dot}~\cite{gansner1990technique}; \textit{dagre}~\cite{dagrejs} and both variants of Sugiyama in \textit{OGDF}~\cite{mutzel2007ogdf}: the barycenter heuristic (``\textit{ogdfb}'') and the median heuristic (``\textit{ogdfm}''). 
    We verified all algorithms, including SPX, produced completely upward drawings.
	We measured drawing area (A), stress (ST), and number of crossings (CR). We also measured height and width separately, but found their behavior to be the same as those for drawing area. The results of the experiment are reported in Table~\ref{table:upwardmetrics}. Each cell indicates the number of times each algorithm had the best value for the metric, with ties being attributed to both algorithms. 


	\begin{table}
	\parbox{.45\linewidth}{
\small
		\centering
		\begin{tabular}{| c | c | c | c | c | c |}
			\hline
			\textbf{} & \textbf{dagre} & \textbf{dot} & \textbf{ogdfb} & \textbf{ogdfm} & \textbf{SPX} \\
			\hline
			ST & 0 & 0 & 0 & 0 & 4 \\
			\hline
			A & 0 & 0 & 0 & 0 & 4 \\
			\hline
			CR & 4 & 4 & 4 & 4 & 4 \\
			\hline
		\end{tabular}
    \caption*{4 binary trees}
		\label{table:upward_all_metrics_first}
		}
\hfill
\parbox{.45\linewidth}{
\small
		\begin{tabular}{| c | c | c | c | c | c |}
			\hline
			\textbf{} & \textbf{dagre} & \textbf{dot} & \textbf{ogdfb} & \textbf{ogdfm} & \textbf{SPX} \\
			\hline
			ST & 0 & 0 & 0 & 0 & 30 \\
			\hline
			A & 0 & 0 & 0 & 0 & 30 \\
			\hline
			CR & 2 & 5 & 8 & 11 & 14 \\
			\hline
		\end{tabular}
		\caption*{30 directed acyclic graphs}
		\label{table:upward_all_metrics_second}
		}
	\caption{The number of times each algorithm had the best metric value for  upward drawings of 4 complete balanced binary trees (left) and 30 DAGs (right).}
	\label{table:upwardmetrics}
	\end{table}
	
	Table~\ref{table:upwardmetrics} shows that SPX consistently produces the best drawings across the metrics, although all other algorithms also produce planar layouts for the complete binary trees.
	However, there is a caveat in the measure of area. We do not impose any resolution to the upwardness of the drawings. The SPX drawings are very small in area compared to those generated by the other algorithms. Imposing a resolution constraint could increase crossings and stress, indicating a post-processing to enforce resolution may be a better option. We experimented with a na\"{i}ve scaling parameter which results in very large area. We leave a more appropriate post-processing algorithm as future work.
	

\section{Conclusions and Future Work}
As some of the drawings in this paper show, optimizing just one layout criterion can result in unreadable drawings. It seems like a natural idea to consider approaches that balance multiple layout criteria. 
SPX is an example of such a graph layout framework that balances the optimization of multiple criteria and achieves quality that is close to one criterion state-of-the-art algorithms. Currently SPX considers stress minimization, crossing minimization, crossing angle maximization, and upwardness. A natural direction for future work is to incorporate additional layout criteria. Our current implementation of SPX relies on a combination of stress minimization and a linear program solver. As a result the algorithm is prohibitively slow for large graphs. Possible ways to speed up the algorithm, such as multi-level computation, are worth exploring.

\bibliography{spx}
\bibliographystyle{splncs04}

\newpage
\section{Appendix}

Further detail about the graphs used for evaluation of the upward criteria is in Appendix~\ref{sec:dags}. We provide further evaluation of SPX to algorithms dedicated to crossing minimization in Appendix~\ref{sec:radermacher}. Comparisons between SPX and algorithms designed for crossing angle maximization using the GD Contest graphs are in Appendix~\ref{sec:gdcontest}.

\subsection{Upwardness Evaluation Graphs}
\label{sec:dags}

	We generated a dataset of four complete balanced binary trees and 30
	directed acyclic graphs (DAGs) to evaluate the quality of upward drawings. The binary trees ranged in depth from 2 through 5.
	
    The DAGs were generated with density 2 and number of vertices ranging from 5 to 24. 
	Each DAG was generated by randomly selecting a pair of vertices, and connecting them by a directed edge only if it did not create a cycle. The process continued until the desired density was reached for the given number of vertices. Once the process ended a connectivity check was performed and the instance was discarded if the graph was not connected.

\subsection{Crossing Minimization Comparison with Radermacher et al.}
\label{sec:radermacher}
Radermacher et al.~\cite{Radermacher18} proposed four algorithms
(described in  Section~\ref{sec:background}) 
that target edge crossing minimization specifically:
EI (Edge Insertion), EP (Edge Placement), VI (Vertex Insertion), and VM (Vertex Movement). They compared their results using a collection of 400 graphs, 100 each in four classes---North, Rome, Community, and Plantri. The North and Rome classes are non-planar subsets of the North and Rome (AT\&T) benchmarks, respectively. The Plantri benchmark contains maximal planar graphs with 64 vertices and the Community graphs are generated to resemble social network with communities. The largest graphs in this collection have 100 nodes.

\begin{figure}[htp]
	\centering
	\begin{subfigure}[b]{0.44\textwidth}
		\includegraphics[width=\textwidth]{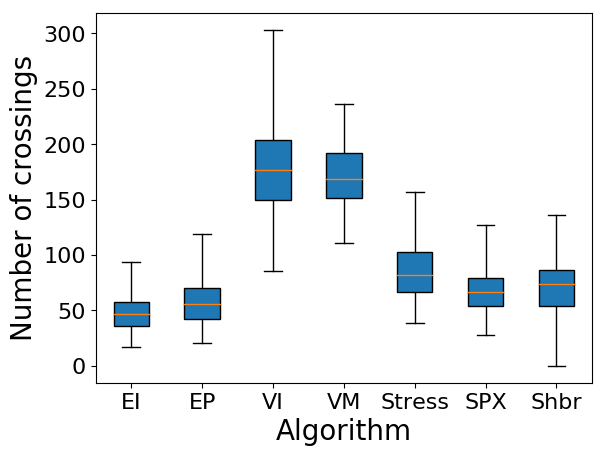}
		\caption{Community graphs.} \label{figure:min_ncr_community}
	\end{subfigure}
	~
	\begin{subfigure}[b]{0.44\textwidth}
		\includegraphics[width=\textwidth]{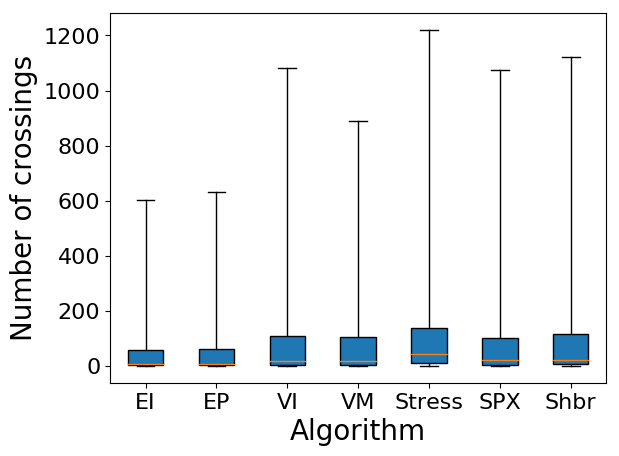}
		\caption{North graphs.} \label{figure:min_ncr_north}
	\end{subfigure}
	\begin{subfigure}[b]{0.44\textwidth}
		\includegraphics[width=\textwidth]{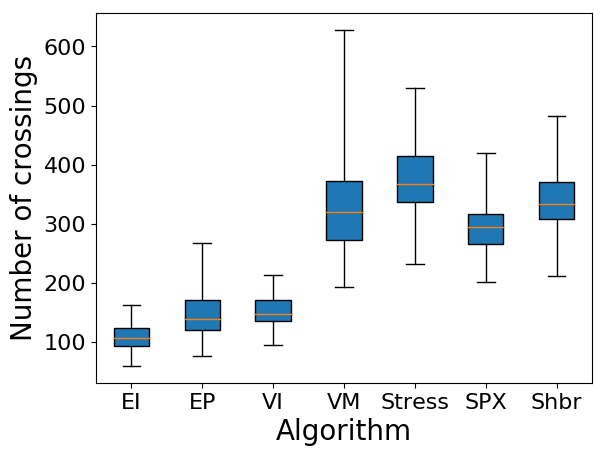}
		\caption{Plantri graphs.} \label{figure:min_ncr_plantri}
	\end{subfigure}
	~
	\begin{subfigure}[b]{0.44\textwidth}
		\includegraphics[width=\textwidth]{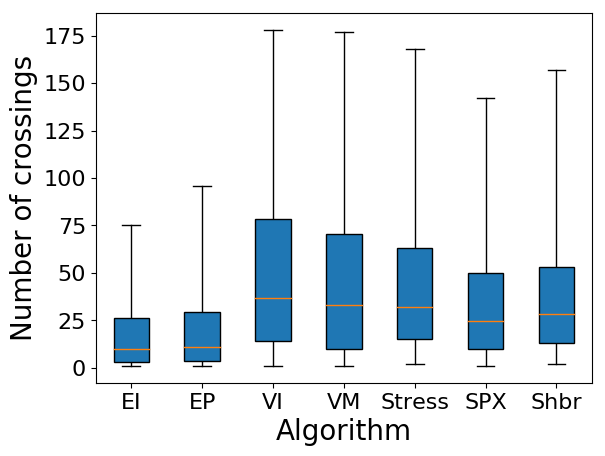}
		\caption{Rome graphs.} \label{figure:min_ncr_rome}
	\end{subfigure}
	\caption{Number of crossings for the 7 algorithms across 400 graphs - 100 each from the 4 classes (Community, North, Plantri, and Rome)}
	\label{fig:min_ncr_100}
\end{figure}

\begin{table}[htp]
\centering
\begin{tabular}{|l|l|l|l|l|l|l|}
\hline
\textbf{Alg} & \textbf{Mean} & \textbf{Min} & \textbf{Q1} & \textbf{Median} & \textbf{Q3} & \textbf{Max} \\
\hline
EI & 48.37 & 17 & 35 & 47 & 58 & 94 \\
\hline
EP & 58.29 & 21 & 42 & 56 & 71 & 119 \\
\hline
VI & 176.78 & 86 & 149 & 177 & 204 & 303 \\
\hline
VM & 170.02 & 111 & 151 & 169 & 194 & 236 \\
\hline
Stress & 85.70 & 39 & 67 & 82 & 104 & 157 \\
\hline
Shabbeer et al. & 75.24 & 34 & 58 & 75 & 87 & 136 \\
\hline
SPX & 67.86 & 28 & 54 & 66 & 79 & 127 \\
\hline
\end{tabular}
\caption{Number of crossings across 100 Community graphs for the 7 algorithms - four variants of Radermacher et al.(EI, EP, VI, VM), stress majorization, Shabbeer et al.'s stress-plus-crossing minimization, and SPX's stress-plus-crossing minimization.}
\label{table:min_ncr_community}
\end{table}

\begin{table}[htp]
\centering
\begin{tabular}{|l|l|l|l|l|l|l|}
\hline
\textbf{Alg} & \textbf{Mean} & \textbf{Min} & \textbf{Q1} & \textbf{Median} & \textbf{Q3} & \textbf{Max} \\
\hline
EI & 47.18 & 1 & 2 & 6 & 61 & 601 \\
\hline
EP & 55.51 & 1 & 2 & 6 & 63 & 630 \\
\hline
VI & 84.39 & 1 & 4 & 20 & 118 & 1080 \\
\hline
VM & 77.99 & 1 & 3 & 17 & 113 & 891 \\
\hline
Stress & 115.69 & 1 & 10 & 44 & 137 & 1220 \\
\hline
Shabbeer et al. & 95.19 & 0 & 6 & 23 & 123 & 1121 \\
\hline
SPX & 91.93 & 1 & 4 & 20 & 114 & 1073 \\
\hline
\end{tabular}
\caption{Number of crossings across 100 North graphs}
\label{table:min_ncr_north}
\end{table}

\begin{table}[htp]
\centering
\begin{tabular}{|l|l|l|l|l|l|l|}
\hline
\textbf{Alg} & \textbf{Mean} & \textbf{Min} & \textbf{Q1} & \textbf{Median} & \textbf{Q3} & \textbf{Max} \\
\hline
EI & 109.45 & 59 & 92 & 107 & 124 & 163 \\
\hline
EP & 144.97 & 76 & 119 & 139 & 170 & 267 \\
\hline
VI & 151.87 & 94 & 135 & 147 & 171 & 213 \\
\hline
VM & 327.39 & 192 & 272 & 320 & 372 & 628 \\
\hline
Stress & 374.65 & 232 & 337 & 367 & 414 & 529 \\
\hline
Shabbeer et al. & 338.53 & 212 & 307 & 334 & 372 & 482 \\
\hline
SPX & 296.54 & 201 & 266 & 294 & 317 & 420 \\
\hline
\end{tabular}
\caption{Number of crossings across 100 Plantri graphs}
\label{table:min_ncr_plantri}
\end{table}

\begin{table}[htp]
\centering
\begin{tabular}{|l|l|l|l|l|l|l|}
\hline
\textbf{Alg} & \textbf{Mean} & \textbf{Min} & \textbf{Q3} & \textbf{Median} & \textbf{Q1} & \textbf{Max} \\
\hline
EI & 16.95 & 1 & 3 & 10 & 27 & 75 \\
\hline
EP & 19.51 & 1 & 4 & 11 & 30 & 96 \\
\hline
VI & 49.59 & 1 & 14 & 37 & 80 & 178 \\
\hline
VM & 47.14 & 1 & 10 & 33 & 71 & 177 \\
\hline
Stress & 44.41 & 2 & 15 & 32 & 63 & 168 \\
\hline
Shabbeer et al. & 38.71 & 2 & 13 & 29 & 53 & 157 \\
\hline
SPX & 33.87 & 1 & 10 & 25 & 50 & 142 \\
\hline
\end{tabular}
\caption{Number of crossings across 100 Rome graphs}
\label{table:min_ncr_rome}
\end{table}

 We report the resulting distribution of crossings in Tables~\ref{table:min_ncr_community} through~\ref{table:min_ncr_rome} and in the boxplots in Fig.~\ref{fig:min_ncr_100}(a - d). The values for stress and the Radermacher et al. algorithms are from Radermacher et al.~\cite{Radermacher18}. Our SPX formulation for edge crossings always performs better than stress minimization alone and Shabbeer et al. across these graphs. It also outperforms VI and VM on the community class of graphs, and VM on the Plantri collections of graphs.

\subsection{Comparison Across Multiple Criteria: GD Contest Graphs}
\label{sec:gdcontest}
We compare our SPX formulation for optimizing stress, edge crossings, and crossing angle as described in Section~\ref{subsec:angle} to two algorithms optimized for crossing angle only. While the SPX formulation cannot outperform the algorithms optimized for a single criteria on that criteria, we also present results for other readability criteria, demonstrating the trade off in optimizing for one criteria versus multiple.

\begin{figure}[htp]
	\centering
	\includegraphics[width=0.75\textwidth]{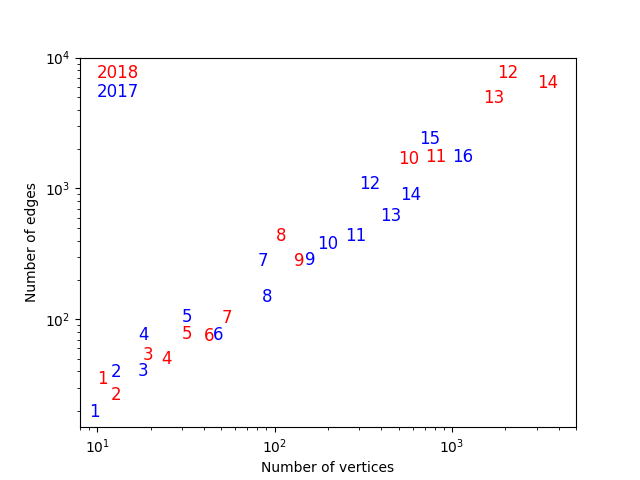}
	\caption{Number of vertices and edges in 2018 and 2017 graphs from Graph Drawing Live Challenge (automatic) - Crossing angle maximization } 
	\label{figure:graph_stats}

\end{figure}

The International Symposium on Graph Drawing and Network Visualization held live challenges~\cite{gdchallenge2018} in 2017 and 2018 with the goal of maximizing the crossing angles in a collection of graphs. Participants were given a set of 16 and 14 graphs, respectively. Figure~\ref{figure:graph_stats} summarizes the graphs in terms of number of vertices and edges. As SPX is computationally demanding, we focus our results on the small and medium size contest graphs (first nine in each set), for a total of 18 graphs.

We compare our SPX formulation to the first and second place teams, Universit\"at T\"ubingen~\cite{Bekos18} and Karlsruhe Institute of Technology (KIT)~\cite{Demel2018AGH} of the 2018 contest~\footnote{In 2017, KIT placed first and T\"ubingen placed second.}. 
We re-ran the algorithms from T\"ubingen and KIT
across the contest graphs to update their 2017 results, to remove the time constraints imposed by the live challenge, and to calculate additional graph layout criteria.
The crossing angle results are summarized in Table~\ref{table:all_teams_2018_2017}.

\begin{table}
	\parbox{.45\linewidth}{
		\centering
		\begin{tabular}{| p {1.1 cm} | p {1.6 cm} | p {1.0 cm} | p {1.0 cm} |}
			\hline
			\textbf{Graph} & \textbf{T\"ubingen} & \textbf{KIT} & \textbf{SPX} \\
			\hline
			1 & 87.71 & 55.96 & 81.18 \\
			\hline
			2 & 89.89 & 89.99 & 89.67 \\
			\hline
			3 & 89.56 & 89.99 & 89.99 \\
			\hline
			4 & 88.90 & 66.45 & 68.43 \\
			\hline
			5 & 78.19 & 53.57 & 32.55 \\
			\hline
			6 & 89.74 & 72.48 & 82.12  \\
			\hline
			7 & 82.81 & 59.69 & 42.81 \\
			\hline
			8 & 42.66 & 17.97 & 3.01 \\
			\hline
			9 & 83.22 & 88.93 & 61.98 \\
			\hline
			
		\end{tabular}
		\caption*{(a) Graphs from 2018}
		\label{table:all_teams}
	}
	\hfill
	\parbox{.45\linewidth}{
		\centering
		\begin{tabular}{| p {1.1 cm} | p {1.6 cm} | p {1.0 cm} | p {1.0 cm} |}
			\hline
			\textbf{Graph} & \textbf{T\"ubingen} & \textbf{KIT} & \textbf{SPX} \\
			\hline
			1 & 89.78 & 77.53 & 89.99 \\
			\hline
			2 & 88.68 & 69.29 & 70.63 \\
			\hline
			3 & 89.94 & 89.98 & 89.99 \\
			\hline
			4 & 89.04 & 52.09 & 41.56 \\
			\hline
			5 & 86.96 & 56.87 & 37.74 \\
			\hline
			6 & 89.72 & 89.90 & 86.84 \\
			\hline
			7 & 61.79 & 42.93 & 20.72 \\
			\hline
			8 & 89.28 & 74.14 & 46.65 \\
			\hline
			9 & 88.20 & 47.64 & 15.46 \\
			\hline
		\end{tabular}
		\caption*{(b) Graphs from 2017}
		\label{table:all_teams_2017}
	}
	\caption{Results from T\"ubingen, KIT, and SPX implementation on crossing angle maximization, Graph Drawing Live Challenge (automatic)}
	\label{table:all_teams_2018_2017}
\end{table}

As expected, the algorithms designed to only optimize crossing angles always perform best on the challenge graphs. However, the SPX formulation performs reasonably well on the small and medium graphs, generally within a factor of two of the other two teams and sometimes comparably. 

We now examine several other readability criteria across the layouts obtained by the different algorithms. In particular, we consider stress, neighborhood preservation, edge crossings, and drawing area measured using the algorithm available from De Luca~\cite{felicemetrics}.

\begin{table}
	\centering
	\begin{tabular}{| c | c | c | c |}
		\hline
		\textbf{Metric} & \textbf{T\"ubingen} & \textbf{KIT} & \textbf{SPX} \\
		\hline
		Crossing angle & 25 & 38 & 45 \\
		\hline
		Stress & 51 & 34 & 23 \\
		\hline
		Neighborhood preservation & 44 & 28 & 26 \\
		\hline
		Drawing area & 36 & 54 & 18 \\
		\hline
		Edge crossings & 30 & 32 & 37 \\
		\hline
	\end{tabular}
	\caption{Comparison among three algorithms with respect to five metrics for the first 9 graphs each (18 graphs total) of 2018 and 2017 challenge (lower values are better). Winning algorithm gets 1 while the losing algorithm gets 3 for each graph. Each row gives the algorithm's score on a particular measurement. }
	\label{table:ang_all_metrics_medium}
\end{table}

Table~\ref{table:ang_all_metrics_medium} summarizes the results for the 18 graphs across all three algorithms. The values are based on a voting system where the algorithm with the best value for a graph gets 1 point, the second gets 2, and the third gets 3. The table shows the total values across all graphs. Lower values are better.

SPX achieves consistently lower values of stress, drawing area, and neighborhood preservation in comparison to the other two algorithms.  Similarly, SPX ranks third for all graphs on crossing angle and edge crossings, but the score is comparable (often within a few degrees) to the second team.

\begin{figure*}[tp]
\centering
\includegraphics[width=1.1\columnwidth]{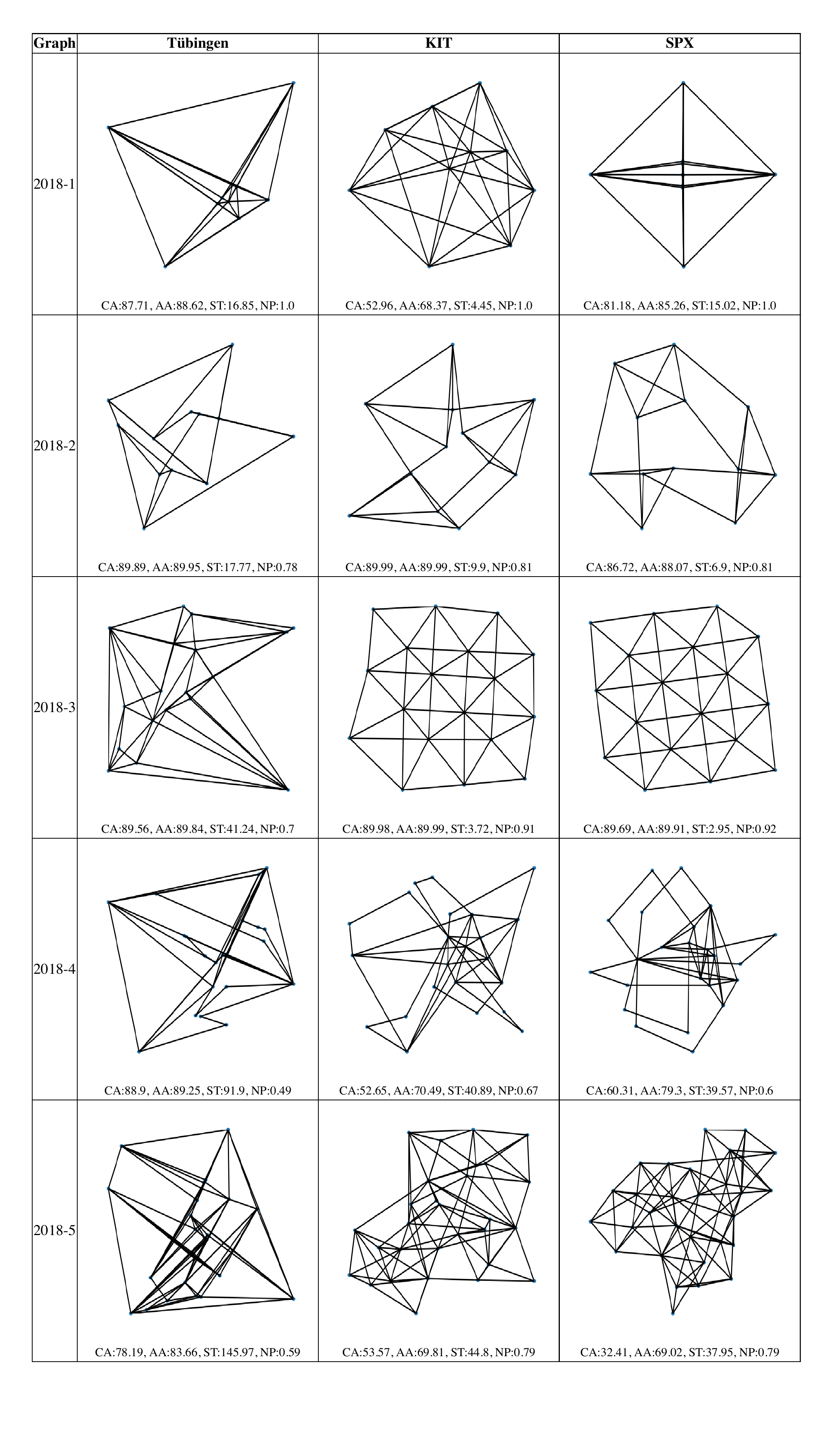}
\caption{Comparison among T\"ubingen, KIT, and SPX algorithm on graphs 1 to 5 from graph drawing challenge 2018 crossing angle maximization} \label{figure:comparison_1_5}
\end{figure*}

\begin{figure*}[tp]
\centering
\includegraphics[width=1.1\columnwidth]{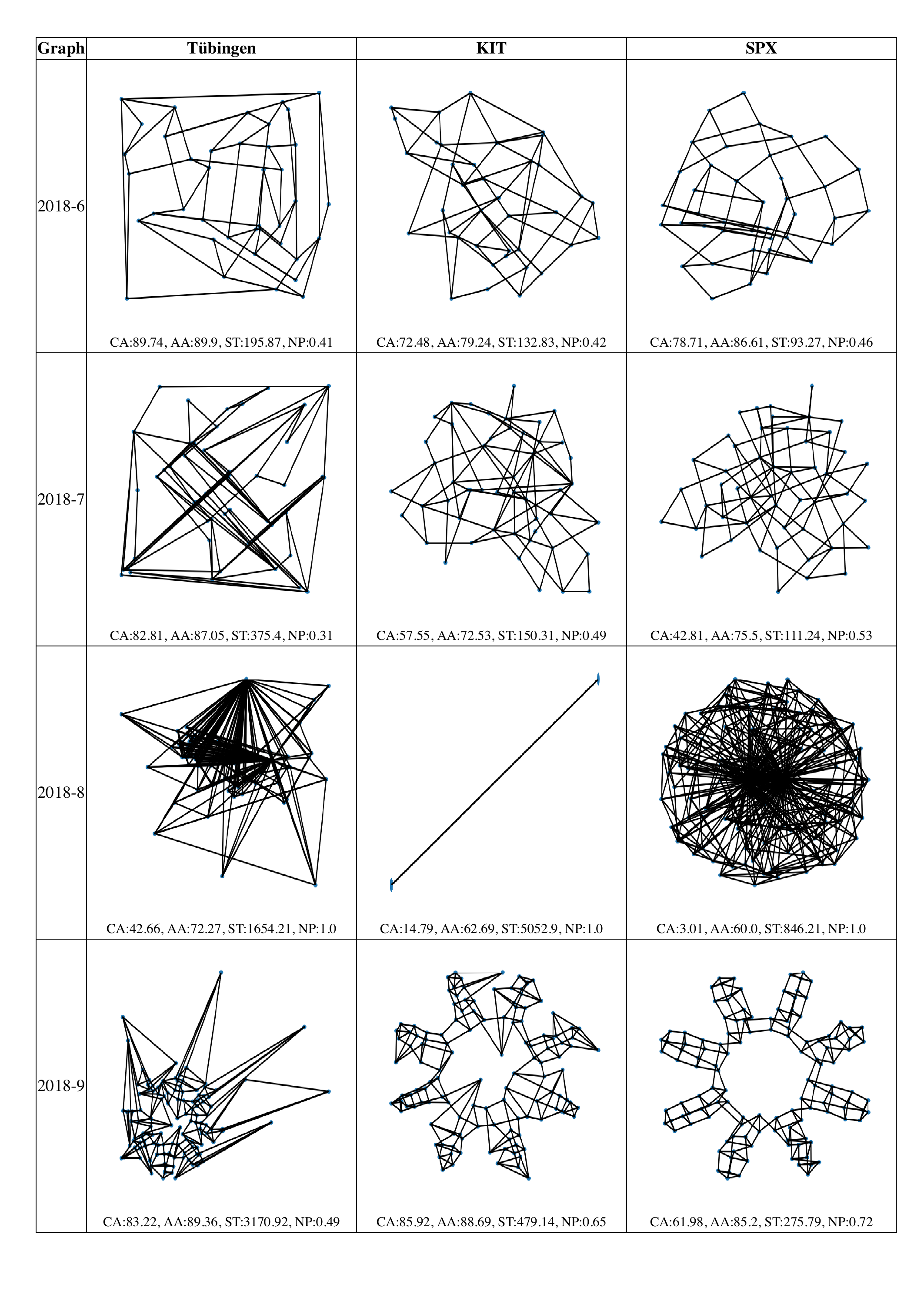}
\caption{Comparison among T\"ubingen, KIT, and SPX algorithm on graphs 6 to 9 from graph drawing challenge 2018 crossing angle maximization} \label{figure:comparison_6_10}
\end{figure*}

\begin{figure*}[tp]
\centering
\includegraphics[width=1.1\columnwidth]{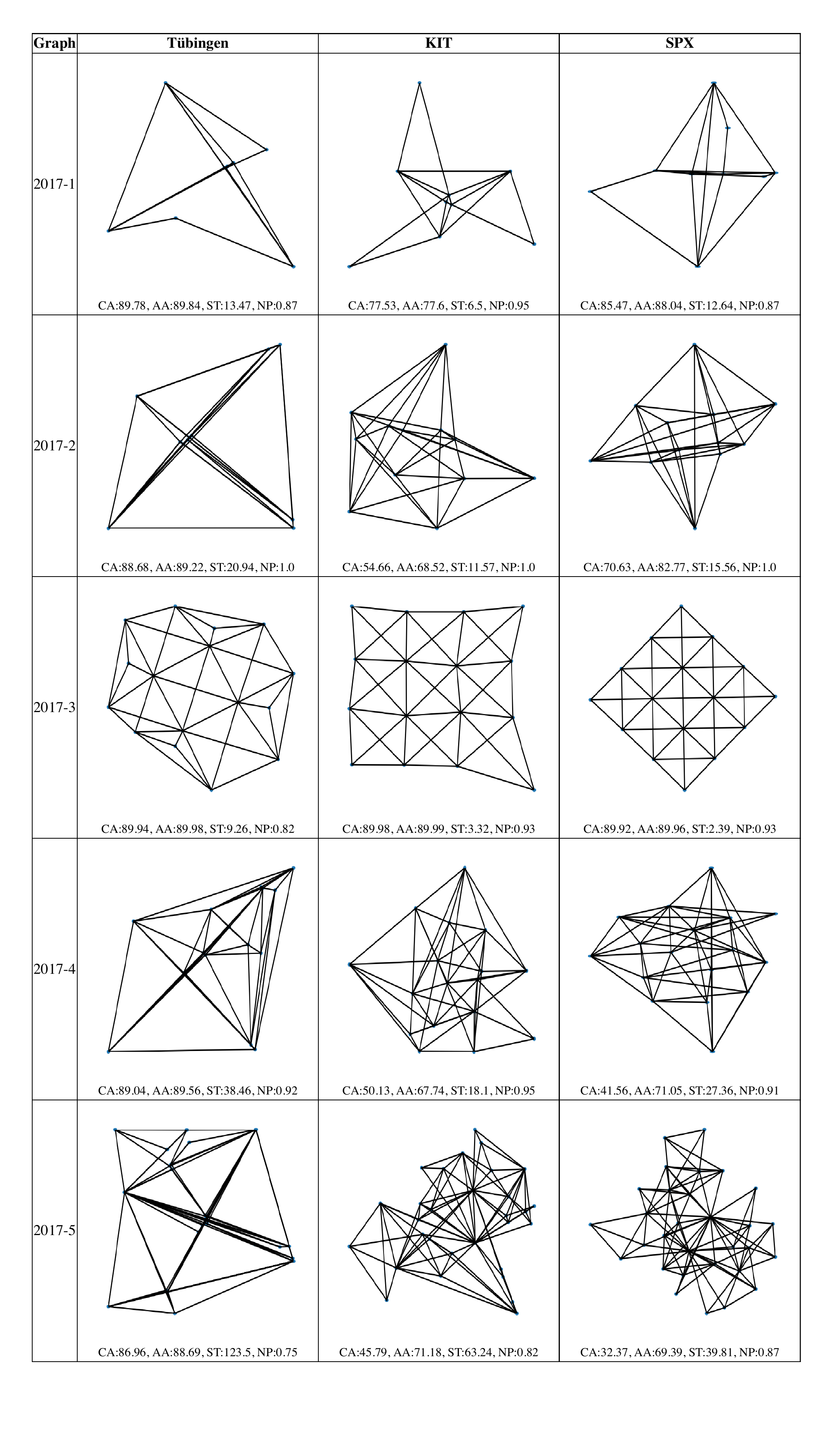}
\caption{Comparison among T\"ubingen, KIT, and SPX algorithm on graphs 1 to 5 from graph drawing challenge 2017 crossing angle maximization} \label{figure:comparison_1_5_17}
\end{figure*}

\begin{figure*}[tp]
\centering
\includegraphics[width=1.1\columnwidth]{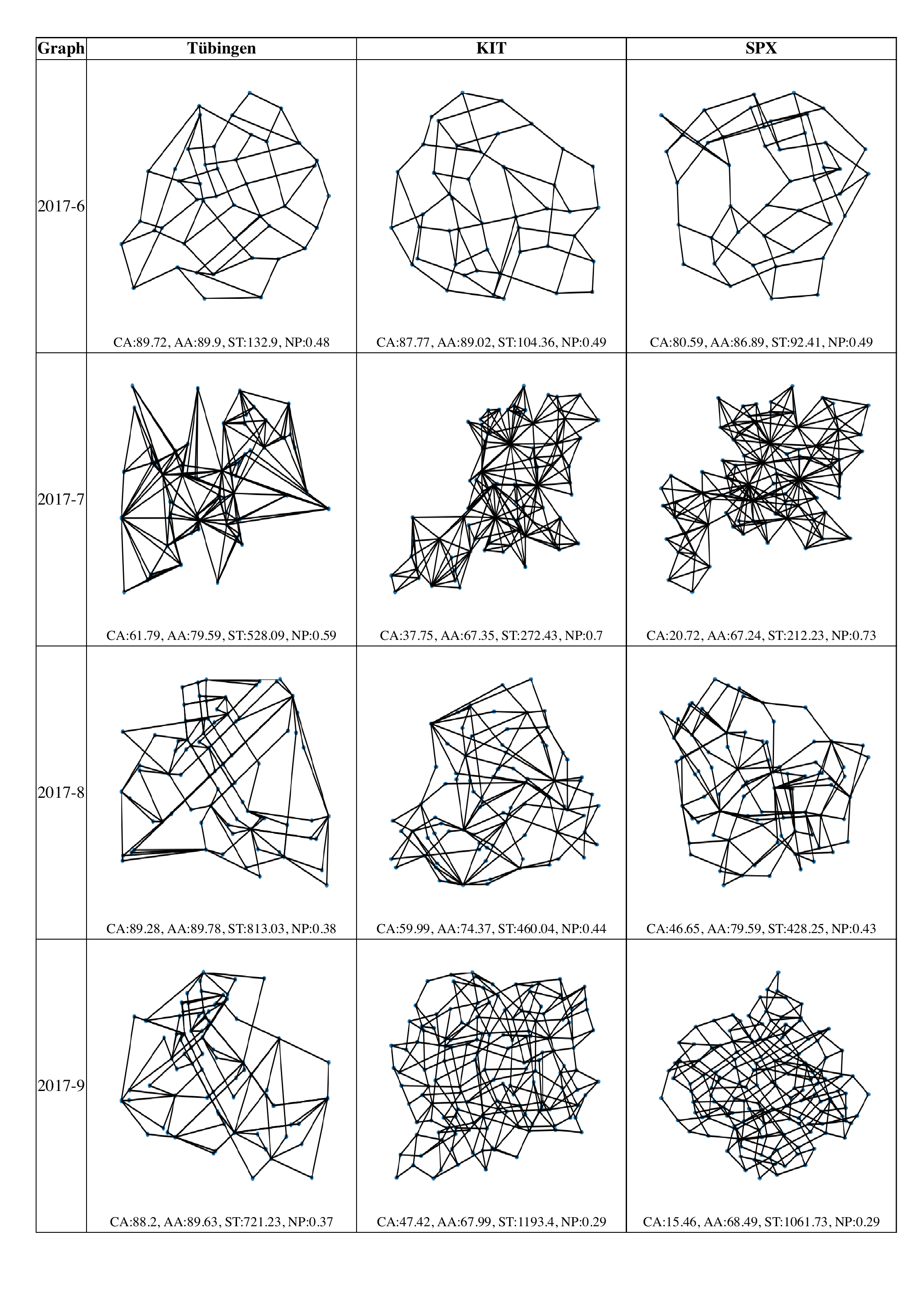}
\caption{Comparison among T\"ubingen, KIT, and SPX algorithm on graphs 6 to 9 from graph drawing challenge 2017 crossing angle maximization} \label{figure:comparison_6_10_17}
\end{figure*}

Figures~\ref{figure:comparison_1_5} through~\ref{figure:comparison_6_10_17} show there are trade-offs in maximizing the graph crossing angle with the overall quality of the layout. 

\subsection{Running Time}
For the 25 community graphs, the minimum, average, and maximum running time for all instances of SPX algorithm is 5.84, 9.72, and 17.58 minutes respectively. The maximum running time of the KIT algorithm is less than a minute. We have run the T\"ubingen algorithm for 10 minutes and taken the best drawing for every instance. The current implementation of SPX has reasonable running time for relatively small graphs. As we mentioned earlier, it finishes in couple of minutes for the graphs of Figure~\ref{figure:community_graph_summary_single}. All of these 25 graphs has 100 vertices and the number of edges varies from 120 to 150. As the size of the graph increases, the running time increases too. For example, consider the 11th graph of 2018 GD contest. It has 709 vertices and 1602 edges. SPX takes around two hours for this graph.

\end{document}